\documentclass{rspublic}

\usepackage{graphicx}
\usepackage{natbib}

\renewcommand{\citep}[1]{(\citeauthor{#1} \citeyear{#1})}

\def\hh{H$_2$}
\def\hhhp{H$_3^+$}
\def\hhdp{H$_2$D$^+$}
\def\ddhp{D$_2$H$^+$}
\def\dddp{D$_3^+$}
\def\hho{H$_2$O}
\def\hhhop{H$_3$O$^+$}
\def\hcop{HCO$^+$}
\def\dcop{DCO$^+$}
\def\nn{N$_2$}
\def\nnhp{N$_2$H$^+$}

\def\zcr{$\zeta_{\rm CR}$}
\def\pow#1#2{#1$\times$10$^{#2}$}
\def\smm{(sub-)\-mil\-li\-me\-tre}
\def\ccm{cm$^{-3}$}
\def\scm{cm$^{-2}$}
\def\gtsim{{_>\atop{^\sim}}}
\def\ltsim{{_<\atop{^\sim}}}

\begin{document}

\title[\hhhp\ and \hhdp\ in star-forming regions]
{Using \hhhp\ and \hhdp\ as probes of star-forming regions}

\author[F.F.S.\ van der Tak]{Floris F.~S.\ van der Tak}

\affiliation{SRON, Landleven 12, 9747 AD Groningen, The Netherlands}

\label{firstpage}

\maketitle

\begin{abstract}{Interstellar medium -- Molecular processes -- Radio
    telescopes -- Astrochemistry}
The \hhhp\ and \hhdp\ ions are important probes of the physical and chemical
conditions in regions of the interstellar medium where new stars are
forming. This paper reviews how observations of these species and of heavier
ions such as \hcop\ and \hhhop\ can be used to derive chemical and kinematic
properties of nearby pre-stellar cores, and the cosmic-ray ionisation rate
toward more distant regions of high-mass star formation. Future
prospects in the field are outlined at the end.

\end{abstract}

\section{Introduction}
\label{sec:intro}

The formation of stars and planets is a crucial step in the life cycle of matter
in galaxies, but remains poorly understood.
While optical and near-infrared observations are useful to locate newborn stars 
and measure the distribution of ionised gas, the cool and dense gas where the
actual star formation takes place is best probed by long-wavelength (infrared
through radio) observations. Molecular spectroscopy is an essential part of this
effort, thanks to the richness of molecular spectra with many rotational and
rovibrational lines which probe a large range of temperatures, densities, and
optical depths. 

This paper reviews the role of line observations of \hhhp\ and its isotopologues
in the determination of basic parameters of star-forming regions. 
After an outline of modern methods to obtain descriptions of the temperature and
density structure of star-forming regions (\S~\ref{s:sfr}), \S~\ref{s:zeta}
shows how observations of \hhhp\ (and other ions) may be used to measure the
ionisation rate of molecular gas. Section~\ref{s:deut} discusses determinations
of the chemistry and kinematics of pre-stellar cores from \hhdp\ and \ddhp\
observations, and \S~\ref{s:future} concludes with an outline of future
opportunities.

\section{Measuring molecular abundances}
\label{s:sfr}

The interpretation of molecular spectra from star-forming regions
requires knowledge of their physical structure.
The envelopes of young stars have temperature and density distributions which
are strongly centrally peaked. For example, in the envelope of AFGL 2136, one of
the first sources where \hhhp\ was discovered, the temperature drops from 450\,K
at \pow{6}{15}\,cm to 22\,K at \pow{2}{18}~cm from the central star, while the
density drops from $\sim$10$^7$ to $\sim$10$^4$\,\ccm\ over the same distance.
%
%
Such gradients lead to significant chemical stratification.  In the cool outer
envelope, the gas-phase chemistry is dominated by ion-molecule reactions which
do not have activation barriers. Closer to the central star, neutral-neutral
reactions become increasingly important.
In addition, at low temperatures, atoms and molecules freeze out on grain
surfaces where they may react with each other. The dominant type of reaction is
hydrogenation at low densities and oxygenation at higher densities. The
resulting ice mantles evaporate off the dust grains when the temperature 
exceeds $\sim$100~K.
%

The temperature and density gradients in protostellar envelopes also manifest
themselves as gradients in the molecular excitation. 
While in the case of infrared observations of rovibrational bands, the
excitation temperature can be measured directly, the determination of molecular
column densities from radio observations often relies on a single or just a few
transitions. 
To be able to compare column densities from different molecules which may have
been derived from lines with very different upper level energies and/or
radiative decay rates (hence critical densities), a comprehensive excitation
model of the source is essential. 

Modern schemes to construct models of star-forming regions are discussed by 
\citet{doty:16293}. Total mass and temperature structure of the envelope is
obtained from dust continuum observations, while the density structure is
inferred from molecular lines. In constraining model parameters, it is important
to be aware that different lines are observed with different beam sizes and thus
do not sample quite the same gas. For further details please consult 
\citet{vdtak:catania}.

\section{Probing the cosmic-ray ionisation rate}
\label{s:zeta}

One key parameter of star-forming regions is the ionisation fraction of the
molecular gas. First, the electron fraction determines the influence of magnetic
fields on the dynamics, in particular the efficiency of magnetic support
against gravitational collapse. Furthermore, the ionisation fraction sets the
time scale of ion-molecule chemistry, which is the most important type of
chemical reactions in the gas phase at temperatures of $\sim$10\,K and
densities of $\sim$10$^4$\,\ccm.
Under these conditions, the bulk of the ionisation is due to `cosmic
rays' ($\sim$100~MeV protons, produced in supernovae), at a rate \zcr,
while ionisation by stellar$\sim$1--10~keV X-rays 
becomes important on small scales \citep{stauber:x-ray}.
To estimate \zcr, \citet{vdtak:zeta} have used models of the
temperature and density structure of the envelopes of seven young
high-mass stars to model observations of \hhhp\ absorption and \hcop\
emission lines. They find \zcr\ $\sim$\pow{3}{-17}\,s$^{-1}$, in good
agreement with measurements of the local flux of low-energy cosmic
rays with the Voyager and Pioneer spacecraft \citep{webber:voyager}.
The models reproduce $\sim$half of the observed $N$(\hhhp), suggesting
that the other half arises in intervening clouds.

%
%

Recent observations suggest strong variations of \zcr\ between lines of sight,
which appear related to column density. On the one hand, observations of \smm\ 
emission lines of \dcop\ and other ions towards the low-mass pre-stellar core
LDN~1544 indicate a decrease of \zcr\ by a factor of $\sim$10
\citep{caselli:zeta}, although depletion of CO and other neutrals onto grains
may influence this result.
On the other hand, observations of \hhhp\ absorption lines toward the
$\zeta$~Per diffuse cloud \citep{mccall:zeta_per} indicate an enhancement of
\zcr\ by a factor of $\sim$10 \citep{lepetit:zeta_per}. 
These estimates are quite reliable now there is a consensus between experiment
and theory on the rate coefficient of dissociative recombination of \hhhp\ at
low temperatures (see Greene, this conference).
Even larger cosmic-ray ionisation rates of several 10$^{-15}$~s$^{-1}$
are found towards the Galactic Centre, where \citet{oka:sgra} obtain
$N$(\hhhp)$\sim$\pow{4}{15}~\scm\ from multi-line \hhhp\ observations 
(see also Geballe, these proceedings).
%
However, this $N$(\hhhp) estimate may change slightly if infrared
pumping is taken into account.
The high stellar density in the Galactic Centre implies an infrared
radiation field which is $\sim$1000 times higher than in local clouds.
Furthermore, the assumption of steady-state excitation of \hhhp\ may
break down at the high \hhhp\ formation and destruction rates in the
Galactic Centre clouds. Calculations of $N$(\hhhp) and \zcr\ toward
Sgr~A which take these effects into account are needed (see also Le
Petit, these proceedings).
Nevertheless, an enhanced ionisation rate in the inner 250\,pc of our
Galaxy is consistent with its strong synchrotron radio emission which
signifies an enhanced electron component of the cosmic rays.
Furthermore, the Galactic Centre is a strong source of X-rays
which may provide additional ionisation.

Even without the Galactic Centre region, evidence is mounting that \zcr\ is
higher in regions with lower column density.
Cosmic rays penetrate to a typical surface density of
$\sim$50\,g\,\scm\ \citep{umebayashi:cosmic-rays}, 
which corresponds to an $N$(\hh) of $\sim$10$^{25}$~\scm.
Even the Sgr~B2 clouds, which are the most massive clouds in the
Galaxy, only have $N$(\hh) $\sim$ \pow{1-3}{24}\,\scm, so that absorption
of cosmic rays probably only plays a minor role except very locally.
Instead, the apparent scaling of the ionisation rate with density may
be due to cosmic-ray \textit{scattering} off plasma waves. This process is
more efficient in denser clouds where magnetic fields are stronger
\citep{padoan:cr_mhd}, which would explain the observed trend. 

\citet{liszt:grains} has suggested that the models for star-forming 
regions fail to include recombination on dust grains.
This process is certainly important for diffuse clouds, where grains
are small and strongly positively charged by UV photons \citep{vdishoeck:diffuse}.
In dense star-forming regions, coagulation increases the
average size of the grains and thus reduces their abundance by number,
given the observed constant mass ratio of dust and gas \citep{ossenkopf:opacities}.
%
%
However, grain charging in these regions is mainly by collisions, 
and the negatively charged dust grains are very good at neutralizing
positive gas-phase ions \citep{tielens:ism}.
%
The processes governing the size and charge of the dust in dense
clouds are thus quite different from those in diffuse clouds, and
detailed models are needed to assess their importance for the overall
ionisation balance.

One critical parameter of chemical models of star-forming regions is
the ratio of the He and H ionisation rates.
Only the He ions are able to break strong chemical bonds such
as the triple bond of the CO molecule, so that the atomic carbon abundance for
instance depends sensitively on $\zeta_{\rm He}$/$\zeta_{\rm H}$.
This ratio is insensitive to the slope of the cosmic-ray energy spectrum
since most damage is done by secondary electrons, whose spectrum
depends on the electron fraction and the H/\hh\ ratio of the gas.
Observational constraints on these quantities are urgently needed.

\section{Chemistry and kinematics of pre-stellar cores}
\label{s:deut}

Lacking an allowed rotational spectrum, the \hhhp\ ion needs a
mid-infrared `background lamp' to be visible in absorption. 
Using \hhdp\ as an
alternative route to \zcr\ is problematic because the strong
temperature dependence of the \hhdp/\hhhp\ ratio means that the two
species do not trace the same gas.
Other ions such as HCO$^+$ and \hhhop\ are more suitable as
substitutes of \hhhp, particularly if observations of CO and 
\hho\ are available \citep{vdtak:h2o}.
Instead, the astrophysical use of \hhdp\ is that of a `chemical filter' to image the
distribution of cold ($\ltsim$10\,K) dense ($\gtsim$10$^5$\,\ccm) gas. Under these
conditions, the formation of \hhdp\ from HD~$+$~\hhhp\ is strongly enhanced
because the back reaction has a barrier; and its destruction is reduced because
the main destroyers of \hhdp, CO, O and \nn, freeze out onto the surfaces of
dust grains. See \citet{roberts:models} and Roberts (these proceedings) for a
detailed description of these processes.

The detection of strong \hhdp\ 372~GHz emission toward the pre-stellar core
LDN~1544 indicates the existence of a region of several 10$^{16}$\,cm radius where all
elements heavier than He are frozen out onto dust grains. Under these
extreme conditions, \hhhp\ and its isotopologues
become the dominant carriers of positive charge, a role normally
played by metals (S$^+$, Fe$^+$) and CNO molecules (HCO$^+$, \hhhop, \nnhp).
Additional evidence for CNO-free zones comes from the detection of \ddhp\
\smm\ emission toward the pre-stellar core LDN~1689N \citep{vastel:d2h+} and
other cores (Vastel, these proceedings). However, observations of deuterated
ammonia toward the same cores indicate that the depletion may not be as complete
as thought before \citep{lis:nd2h}, or limited to a smaller region. Indeed,
the \hhdp\ abundance of $\sim$\pow{1}{-9} in LDN~1544 by \citet{caselli:h2d+} 
has probably been overestimated by up to a factor of 10 through the collisional
rate coefficient \citep{vdtak:l1544} and the kinetic temperature (Crapsi et al.,
in prep.). Nevertheless, the \hhdp\ abundance is clearly higher in pre-stellar
cores than around low-mass protostars (\citealt{stark:h2d+}, \citeyear{stark:16293}).

Thus, although better data are needed to determine the sizes of the CNO-free 
zones, the very existence of such zones at the centres of pre-stellar cores
implies that the relative abundances of \hhhp, \hhdp\ and \ddhp\ are sensitive to 
the grain size \citep{flower:grain_size} and the o/p \hh\ ratio \citep{flower:op-h2}.
Normally, these parameters are very hard to constrain by observations because of
other competing effects. The centres of pre-stellar cores thus form a unique
laboratory to study basic physical parameters of interstellar gas.

In most radio observations, the lines are spectrally resolved, and the line profiles 
can be used to extract additional information about the source.
A prime example is the double-peaked line profile of \hhdp\ toward the
pre-stellar core LDN~1544 \citep{vdtak:l1544}.
The central absorption dip implies a gradient in the excitation of
\hhdp, which means an increasing density and a constant gas temperature.
Previous models with a constant density at the centre and a centrally
decreasing gas temperature are ruled out by the new \hhdp\ observations.
Furthermore, the narrowness of the absorption from the very light
\hhdp\ molecule implies that radial motions are subthermal in the
outer parts of the core but increase inward.
Maps of \hhdp\ and \ddhp\ emission will provide further constraints on
the structure and kinematics of pre-stellar cores (Vastel, these proceedings).

\section{Future directions}
\label{s:future}

The next 5--10 years will bring several new facilities for \smm\ observations of
\hhdp\ and \ddhp. 
Specifically, the APEX telescope can observe the \hhdp\ 372~GHz line about ten
times faster than the CSO (Belloche et al., in prep.). Its recently commissioned
CONDOR receiver for Terahertz frequencies enables us to observe the 1370~GHz
line (Emprechtinger \& Wiedner, this conference). The \ddhp\ molecule may be
observed through its 692~GHz line with the CHAMP heterodyne array receiver,
while its 1477~GHz line can be reached with CONDOR.
Ultimately, the ALMA interferometer will provide sensitive high-resolution
images of the \hhdp\ 372 GHz and \ddhp\ 692~GHz lines.

At mid-infrared wavelengths, higher spectral resolution observations (such as
offered by CRIRES on the VLT) will be crucial to obtain velocity-resolved
profiles of the \hhhp\ absorption. Velocity information is crucial to
distinguish intrinsic absorption from intervening clouds, and will improve the
accuracy of estimates of the cosmic-ray ionisation rate.
Observations of \hhdp, \ddhp\ and \dddp\ mid-infrared absorption lines would allow much
more direct probes of the \hhdp/\hhhp\ ratio. First attempts were unsuccessful
but did result in the serendipitous infrared detection of formaldehyde \citep{roueff:h2co}.

Theoretical calculation of the rate coefficient of de-excitation of \hhdp\ in
collisions with \hh\ would greatly improve the accuracy of
estimates of the \hhdp\ abundances in pre-stellar cores and star-forming regions.
Such calculations are challenging because of the reactive nature of the
interaction, but the He-\hhdp\ system may be a useful starting point.

In summary, \hhhp\ is great, but using it as an astrophysical tool requires
other molecules too. The future will bring excellent observing facilities to
carry out such observations.

\begin{acknowledgements}
   The author thanks Takeshi Oka for the invitation to this meeting,
   and John Black, Carsten Dominik, Eric Herbst, Harvey Liszt and
   Malcolm Walmsley for fruitful discussions.
\end{acknowledgements}

\bibliographystyle{floris}
\bibliography{london}

\end{document}